\documentclass[prb,superscriptaddress,twocolumn,showpacs]{revtex4}
\usepackage{graphicx}
\usepackage{amsmath}

\newcommand{\Fref}[1]{Fig.~\ref{#1}}

\newcommand{\PRL}{Phys. Rev. Lett.}
\newcommand{\PRB}{Phys. Rev. B}
\newcommand{\PC}{Physica C}
\begin{document}

\title{Comment on ``Anisotropic $s$-wave superconductivity in MgB$_2$"}

\author{Todor M. Mishonov}
\email[E-mail: ]{todor.mishonov@fys.kuleuven.ac.be}
\affiliation{Laboratorium voor Vaste-Stoffysica en Magnetisme,
     Katholieke Universiteit Leuven,\\  Celestijnenlaan 200 D, B-3001
 Leuven, Belgium}
\affiliation{Department of Theoretical Physics, Faculty of Physics,
 Sofia University ``St. Kliment Ohridski'', 5 J. Bourchier Blvd.,
 Bg-1164 Sofia, Bulgaria}
\author{Evgeni S. Penev}
\affiliation{Department of Theoretical Physics, Faculty of Physics,
 Sofia University ``St. Kliment Ohridski'', 5 J. Bourchier Blvd.,
 Bg-1164 Sofia, Bulgaria}
\author{Joseph O. Indekeu}
 \email[E-mail: ]{joseph.indekeu@fys.kuleuven.ac.be}
 \affiliation{Laboratorium voor Vaste-Stoffysica en Magnetisme,
      Katholieke Universiteit Leuven,\\  Celestijnenlaan 200 D, B-3001
 Leuven, Belgium}

\begin{abstract}
An analytical result for renormalization of the jump of the heat
capacity $\Delta C/C_N$ by the anisotropy of the order parameter is
derived in the framework of the model proposed by Haas and Maki
[Phys. Rev.~B \textbf{65}, 020502(R) (2001)], for both prolate and
oblate anisotropy. The graph of $\Delta C/C_N$ versus the ratio of the
gaps on the equator and the pole, $\Delta_e/\Delta_p$, of the Fermi
surface allows a direct determination of the gap anisotropy parameter
$\Delta_e/\Delta_p$ using data from specific heat measurements.
\end{abstract}


\maketitle

In a recent Rapid Communication, Haas and Maki~\cite{Haas:01} discuss
a model for the gap anisotropy in MgB$_2$, a material which has
attracted a lot of attention from condensed matter physicists in the
past two years.  A central issue in this work is to propose an
analytic model for analyzing the thermodynamic behavior. Assuming a
spherical Fermi surface a simple gap anisotropy function is suggested,
$\Delta ({\bf k}) \propto 1+ a z^2$, where $z= \cos\theta$, and
$\theta$ is the polar angle. This angular dependence is a natural
linear combination of spherical harmonics (Legendre polynomials
$P_0(z)$ and $P_2(z)$), and can therefore be interpreted in terms of
an s+d~model.
\begin{figure}[h]
\includegraphics[width=0.9\columnwidth]{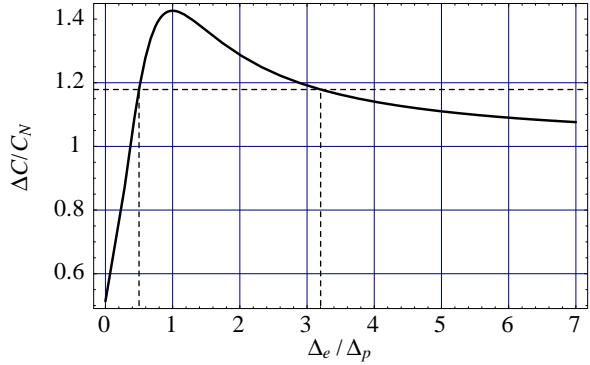}
\caption{Jump of the specific heat $\Delta C/C_N$ as function of ratio
between ``equatorial'' and ``polar'' order parameter
$\Delta_e/\Delta_p$. For one and the same jump $\Delta C/C_N$ we have
prolate and oblate solution.
\label{fig:1} }
\end{figure}

This model leads to useful results for the temperature dependence of
the upper critical field $H_{c2}$ and of the specific heat, which can
be fitted to the experimental data, thereby determining the optimal
anisotropy parameter $a$. Note that $a = \Delta_p/\Delta_e -1$, with
$\Delta_p = \Delta (z=1)$ and $\Delta_e = \Delta (z=0)$.  For example,
the dimensionless ratio $\Delta C/ C_N$ of the specific heat jump at
$T_c$ relative to the normal state specific heat, measured by Wang
\textit{et al.},~\cite{Wang:01} is a tool for estimating $a$. Haas and
Maki~\cite{Haas:01} focused on the case $a>0$.

For oblate-shaped gap anisotropy ($\Delta_e > \Delta_p$), indicated by
recent experiments on the critical field
anisotropy,\cite{Xu:01,Angst:02} we take $a<0$. The sign of $a$ is
relevant for this physical property, but irrelevant for the jump of
the heat capacity. Note that for $a<-1$ the model gives a gapless
superconductor, which is not applicable for MgB$_2$. Following the
weak-coupling BCS approach by the authors~\cite{Haas:01} we derived
the analytic expression
\begin{align}
\frac{\Delta C}{C_N}
  & = \frac{12}{7\zeta(3)}
      \frac{1+ 4 a /3 + 38 a^2/45 + 4 a^3/15 + a^4 /25}
      {1+ 4 a/3 + 6 a^2/5+ 4 a^3/7 + a^4/9},\nonumber \\
a & =\frac{1}{\Delta_e/\Delta_p}-1,\qquad \Delta_e/\Delta_p=\frac{1}{1+a}
\end{align}
valid for arbitrary $a$. This expression can be used in experimental
data analysis and may lead to \emph{two} solutions (oblate, $a<0$;
prolate, $a>0$), for a given specific heat jump.  The relevant example
is shown in \Fref{fig:1}.

For the jump $\Delta C/C_N \approx 1.18$ considered by the
authors,~\cite{Haas:01} we recall $a = 1$ and contribute $a = -
0.6877$. The corresponding anisotropies are $\Delta_e/\Delta_p = 1/2$
(prolate) and $\Delta_e/\Delta_p = 3.202 > 1$ (oblate solution).

This paper was supported by the Flemish programmes VIS, IUAP and GOA.

\end{document}